\documentclass[11pt]{article}

\usepackage{amsmath}  
\usepackage{latexsym}
\usepackage{amssymb}
\usepackage{mathrsfs}

\usepackage{color}

\newcommand{\nonu}{\nonumber \\} 

\newcommand{\hs}[1]{\hspace{#1 mm}}
\newcommand{\beq}{\begin{equation}}
\newcommand{\eeq}{\end{equation}}
\newcommand{\ben}{\begin{eqnarray}}
\newcommand{\een}{\end{eqnarray}}

      \def\cI{{\cal I}}

\def\fa{{\mathfrak a}}
\def\fb{{\mathfrak b}}
\def\fc{{\mathfrak c}}

\def\ca{{\alpha}}
\def\cb{{\beta}}
\def\cc{{\gamma}}
\def\cd{{\delta}}
\def\ck{{\kappa}}

\newcommand{\BB}{\mbox{${\mathbb B}$}}

\newcommand{\mA}{\mathscr{A}}
\newcommand{\mB}{\mathscr{B}}
\newcommand{\mC}{\mathscr{C}}
\newcommand{\mD}{\mathscr{D}}

\def\ft{{ t}}

\newcommand{\mb}[1]{\hs{4}\mbox{#1}\hs{4}}

\def\y2{$Y(\mathfrak{sl}_2)$}
\def\c{\mbox{coef}}

\begin{document}

\ \hfill{LAPTH-042/12}\qquad\qquad\qquad\\[1.2ex]
\begin{center}
\textbf{\Large Algebraic Bethe ansatz for open XXX model\\[1.2ex]
 with triangular boundary matrices}\\[2ex]
\large{S. Belliard$^a$, N. Cramp\'e$^a$ and E. Ragoucy$^b$
\footnote{emails: samuel.belliard@univ-montp2.fr, nicolas.crampe@univ-montp2.fr, ragoucy@lapp.in2p3.fr.}}\\[2.1ex]
$^a$ Laboratoire Charles Coulomb L2C, UMR 
5221,\\
CNRS et Universit\'e Montpellier 2, F-34095 Montpellier, France \\[1.2ex]
$^b$ Laboratoire d'Annecy-le-Vieux de Physique Th{\'e}orique LAPTH,\\
CNRS et Universit{\'e} de Savoie, UMR 5108\\
B.P. 110, F-74941 Annecy-le-Vieux Cedex, France
\\[2ex]
 MSC numbers (2010): 82B23 ; 81R12\\
Keywords: Algebraic Bethe ansatz; integrable spin chain; boundary conditions\\
 \end{center}

\abstract{\textsl{We  consider an open XXX spin chain with two general boundary matrices whose entries obey a  relation, which
is equivalent to the possibility to put simultaneously the two matrices in a upper-triangular form. 
We construct Bethe vectors by means of a generalized algebraic Bethe ansatz. As usual, the method uses 
 Bethe equations and provides transfer matrix eigenvalues.
}}

\null

\medskip

The XXX spin chain \cite{heisen} is one of the most studied integrable model.
When the boundary conditions are periodic, almost everything is known about it, 
starting from its spectrum up to 
asymptotics of correlation functions, and all the possible methods 
(\textit{e.g.} coordinate \cite{bethe} or algebraic \cite{kusk,tafa,KoIzBo,fafa} Bethe ansatz,...) 
have been successfully used. 
One should, however, note that, as for most of quantum integrable models treated with any type of Bethe ansatz, the question of completeness and simplicity of the spectrum remained unsolved for a long time until \cite{MTV}. 

When the model has open boundary conditions, the situation changes drastically. 
Indeed,  the model is known to be integrable for general boundary matrices \cite{sklyanin}, but
for a long time only the case of diagonal boundary matrices was well understood, using  coordinate \cite{gaudin,xxz} or algebraic \cite{sklyanin} Bethe ansatz. Recently \cite{MMR,CR}, 
a first step toward a resolution of the model, when one boundary matrix is triangular and the other one diagonal, was done,   
in the framework of algebraic or coordinate Bethe ansatz.
In order to tackle this problem, the separation of variables has been also used \cite{FSW}. 
The aim of the present letter was to construct an algebraic Bethe ansatz for the open XXX spin chain with two general boundary matrices with one relation among their entries.
This relation is equivalent to the possibility to set simultaneously the two matrices in a upper triangular form\footnote{Of course, the method also works when the two matrices can be simultaneously put in a lower triangular form.}. 
  
When the two boundaries are triangular, the main difficulty lies in the fact that the total longitudinal spin is not preserved anymore. 
To overcome this problem, we allow  the ansatz to have states with  unfixed total longitudinal spin. The same idea was applied successfully in \cite{CR,CRS1,CRS2} in the context of  coordinate Bethe ansatz.
 
\section{Reflection equation and transfer matrix}

The open XXX spin chain Hamiltonian can be constructed from the rational $R$-matrix, satisfying the Yang--Baxter equation \cite{yang,baxter,baxter2,baxter3},
and $K$-matrices, satisfying the reflection equation \cite{chered, sklyanin}. 
In this section, we recall the main step of this construction and give the constraints on the parameters to get triangular boundary matrices.
We  also present its pseudo-vacuum state.

\subsection{Reflection equation}

The $R$-matrix associated with the XXX spin chain has the well-known form
\ben
R(u) &=& \left(\begin{array}{cccc} \fa(u) &0 &0 &0 \\
0 & \fb(u) & \fc &0 \\
0 & \fc & \fb(u) &0 \\
0 &0 &0 & \fa(u)
\end{array}\right)\,,
\een
where 
\begin{eqnarray}
 &&\fa(u)= u+\eta
\mb{;} \fb(u)= u 
\mb{and}
\fc = \eta \;.
\label{eq:abc-yang1}
\end{eqnarray}
This $R$-matrix, associated with the Yangian \y2, satisfies the Yang--Baxter equation \cite{yang,baxter,baxter2,baxter3}
 \ben
&& R_{12}(u-v)R_{13}(u-w)R_{23}(v-w)\ =\ R_{23}(v-w)R_{13}(u-w)R_{12}(u-v)\;.
 \een
As usual, the subscripts of the $R$-matrix indicate the spaces where it acts non-trivially.
This $R$-matrix allows one to construct the monodromy matrix
\begin{equation}\label{eq:T}
 T_a(u)=R_{a1}(u-\xi_1)\ R_{a2}(u-\xi_2)\dots R_{aL}(u-\xi_L)\;,
\end{equation}
where $L$ is the number of sites and $\xi_j$ are free parameters, called inhomogeneity parameters.
This monodromy matrix is the cornerstone of the study of integrable periodic spin chains.
   
To construct integrable spin chains with boundaries, we follow the method introduced in \cite{chered,sklyanin} 
based on the reflection equations
\begin{eqnarray}
\label{RA}
  R_{12}(u-v)\ B_{1}(u)\ R_{12}(u+v)\ B_{2}(u)\!\!\!
&=&\!\!\!B_{2}(u)\ R_{12}(u+v)\ B_{1}(u)\ R_{12}(u-v)\,, \\
\label{RAd}
R_{12}(-u+v)\, \overline{B}_1^t(u)\, R_{12}(-u-v-2\,\eta)\, \overline{B}_2^t(v)\!\!\!
&=&\!\!\!\overline{B}_2^t(v)\, R_{12}(-u-v-2\,\eta)\,  \overline{B}_1^t(u)\, R_{12}(-u+v)\,,\
\end{eqnarray}
where $(\cdot)^t$ stands for the transposition.
For later convenience, we introduce the following operators:
\begin{equation}
 \mA(u)=B_{11}(u)\ ,\quad\mB(u)=B_{12}(u)\ ,\quad\mC(u)=B_{21}(u)\ ,\quad\mD(u)=B_{22}(u)-\frac{\eta}{2u+\eta}B_{11}(u)\,,
\end{equation}
where $B_{ij}(u)$ are the entries of the matrix $B(u)$. The reflection equation (\ref{RA}) provides commutation relations 
between these operators. This computation is well known \cite{sklyanin} (see also \cite{OBA}) and we report a list of these commutation 
relations in  Appendix \ref{App:cr}.

To construct an open spin chain, we need scalar solutions of these reflection equations.
In the  \y2 case, the most general scalar solutions  were classified in \cite{Kmat}. They are given, respectively, for
(\ref{RA}) and (\ref{RAd}), by
\begin{eqnarray}\label{eq:K}
K(u)=
\left(\begin{array}{cc}
	u \,\cb + \ca & u \, \cc \\
	u\,\cd      & -u \,\cb+ \ca
\end{array}\right)\mb{and}
\overline{K}(u)=
\left(\begin{array}{cc}
	(-u-\eta) \,\bar{\cb} + \bar{\ca} & (-u-\eta) \, \bar{\cc} \\
	(-u-\eta)\,\bar{\cd}      & (u+\eta) \,\bar{\cb}+ \bar{\ca}
\end{array}\right)\;,
\end{eqnarray}
where $\ca,\cb,\cc,\cd$ and $\bar{\ca},\bar{\cb},\bar{\cc},\bar{\cd}$ are free parameters.
Using the monodromy matrix $T(u)$ and the scalar solution $K(u)$, 
we construct another solution of the reflection equation (\ref{RA}) via the dressing procedure
\ben\label{eq:B}
B_a(u)=T_a(u)\ K_a(u)\ T^{-1}_a(-u)\;.
\een
We are now in position to introduce the transfer matrix associated with open XXX spin chain
\ben
\ft(u)=tr_a\left(\ \overline{K}_a(u)\ B_a(u)\ \right)\;,
\een
which commutes for different spectral parameters (i.e. $[\ft(u),\ft(v)]=0$).
Finally, the integrable open XXX spin chain Hamiltonian is given by (in the case where $\xi_j=0$)
\ben
H_{XXX open}=\frac{\eta^{2L-1}}{8\alpha\bar\alpha}\frac{d}{du} \ft(u) |_{u=0}=\sum_{j=1}^{L-1} P_{j,j+1}
+\frac{1}{2\bar\alpha} \overline K_1(0)+\frac{1}{4\eta\alpha} K'_L(0)\,,
\een
 where $'$ denotes the derivation and $P_{j,j+1}$ is the permutation operator in spaces $j$ and $j+1$, \textit{e.g.}
\ben
P=\left(\begin{array}{cccc} 1 & 0& 0& 0\\
0 & 0 & 1 &0 \\
0 & 1 & 0 &0 \\
0 &0 &0 & 1
\end{array}\right)\,.
\een

\subsection{Triangularization and constraints on the boundary parameters}

Although the spin chain with boundaries parameterized by the $K$-matrices (\ref{eq:K}) is related to a transfer matrix
(which provides $L$ independent conserved charges), the construction of its eigenvectors remains an open problem. 
The main difficulty
lies in the construction of a pseudo-vacuum, i.e. the determination of one particular Hamiltonian eigenvector.
The case with diagonal boundaries is simpler and has been already treated in \cite{sklyanin}.
Then, using the $R$-matrix invariance
$M_1M_2R_{12}(u)=R_{12}(u)M_1M_2$, valid for any 2 by 2 matrix $M$,
 the case where $K$ and $\overline{K}$ can be simultaneously
diagonalized has been treated in \cite{Gb, GM}.
The most general case treated up to now is when there exists a basis where $K$ is triangular and $\overline{K}$ is diagonal \cite{MMR}.
In this paper, we propose a generalization of the algebraic Bethe ansatz (based on the ideas of  
\cite{CRS1,CR,CRS2}) 
to deal with the case where both $K$-matrices are triangular.

Obviously \cite{Gb, GM, MMR}, if we conjugate both $K$-matrices by a constant matrix, the transfer matrix
eigenvalues are unchanged.
Therefore, we want to find a 2 by 2 matrix $M$ such as $M^{-1}K(u)M$ and $M^{-1} \overline{K}(u) M$ are upper triangular. Unfortunately, it is not always possible.
It is a simple algebra exercise to show that one can do it if and only if the following constraint is valid:
\begin{equation}\label{eq:c}
 (\bar{\cd} \cc - \cd \bar{\cc} )^2 -4 (\cb\bar{\cc}-\bar{\cb} \cc )(\bar{\cd}\cb-\cd\bar{\cb} )=0\;.
\end{equation}
To our knowledge, it is the less restrictive constraint on the boundary parameters for which the eigenvalues and the eigenvectors 
of the transfer matrix are known (see Section \ref{sec:EE}).  Let us stress that the calculation will be done even if the boundary matrices are \textit{not} diagonalizable (\textit{e.g.} when $\beta^2+\gamma\delta=0$ or $\bar\beta^2+\bar\gamma\bar\delta=0$).
Clearly, we can deal with lower triangular matrices in the same way.

\subsection{Pseudo-vacuum}

From now on, 
we assume that the constraint (\ref{eq:c}) is satisfied. 
Then, we can triangularize the $K$-matrices (\ref{eq:K}) to
 get 
\begin{eqnarray}\label{eq:Kt}
K(u)=
\left(\begin{array}{cc}
	u \,b + a & u \, c \\
	0      & -u \,b+ a
\end{array}\right)\mb{and}
\overline{K}(u)=
\left(\begin{array}{cc}
	-(u+\eta) \,\bar{b} + \bar{a} & -(u+\eta) \, \bar{c} \\
	0    & (u+\eta) \,\bar{b}+ \bar{a}
\end{array}\right),
\end{eqnarray}
where $a,b,c$ and $\bar{a},\bar{b},\bar{c}$ are still free parameters. 
The relations between  $a,b,c,\bar{a},\bar{b}$ and $\bar{c}$ and the original parameters are given by
\begin{eqnarray}
&& a=\alpha\,,\quad b^2=\beta^2+\gamma\delta\,,\quad  c=\gamma+\delta 
\end{eqnarray}
and similar relations with ``bar'' parameters.  One possible matrix $M$ relating (\ref{eq:K}) and (\ref{eq:Kt}) is 
\ben
M=\left(\begin{array}{cc}
b+\beta & \delta \\ \delta & b+\beta\end{array}\right)\,.
\een

The transfer matrix to be diagonalized now reads
\begin{equation}\label{eq:tt}
 \ft(u)=\ck_1(u)\ \mA(u)+\ck_2(u)\ \mD(u)
 +\ck_{12}(u) \ \mC(u)\,,
\end{equation}
where we used the notations
\begin{equation}\label{eq:kappa}
 \ck_1(u)=\frac{2(u+\eta)}{2u+\eta}\,(\bar{a}-\bar{b}u)\mb{,}
\ck_2(u)=(u+\eta) \,\bar{b}+ \bar{a}\mb{,}
\ck_{12}(u)=-(u+\eta)\, \bar{c}\,.
\end{equation}

An important point is that for two triangular matrices, there still exists
a simple eigenvector, called pseudo-vacuum. Indeed, let us consider the vector $|\Omega\rangle$ with $L$ spin up, i.e.
\begin{equation}
|\Omega\rangle=\left(\begin{array}{c}
                         1\\0
                        \end{array}\right)
\otimes\left(\begin{array}{c}
                         1\\0
                        \end{array}\right)
\otimes\dots\otimes
\left(\begin{array}{c}
                         1\\0
                        \end{array}\right)\;.
\end{equation}
As explained in \cite{MMR}, when 
we choose the $K$-matrices (\ref{eq:Kt}),
this vector has the following properties:
\begin{eqnarray}
\mA(u)|\Omega\rangle &=& \Lambda_1(u)\, |\Omega\rangle
\mb{with} \Lambda_1(u)\ =\
 (a+bu) \prod_{j=1}^L \frac{\fa(u-\xi_j)}{\fa(-u-\xi_j)}\;;\\
\mD(u) |\Omega\rangle &=& \Lambda_2(u)\, |\Omega\rangle
\mb{with} \Lambda_2(u)\ =\ \frac{2u(a-b(u+\eta))}{2u+\eta}
  \prod_{j=1}^L 
\frac{\fb(u+\xi_j)\fb(u-\xi_j)}{\fa(u+\xi_j)\fa(-u-\xi_j)}\,;\quad\\
\mC(u) |\Omega\rangle &=& 0\;.
\end{eqnarray}
Using these properties, it is easy to show that $|\Omega\rangle$ is an eigenvector of $\ft(u)$
\begin{equation}
 \ft(u)|\Omega\rangle=(\ck_1(u)\Lambda_1(u)
+\ck_2(u) \Lambda_2(u))\ |\Omega\rangle.
\end{equation}

\section{Algebraic Bethe ansatz}

\subsection{Bethe vectors}

We are now in position to propose an ansatz for all the eigenvectors. Let us remark that 
if $\bar c$ in (\ref{eq:Kt}) vanishes, the ansatz used to study the case with both boundaries diagonal is still working.
This trick was used previously in \cite{MMR}. For the case with non-vanishing $\bar c$, we need 
to generalize the ansatz. To this aim, we borrow the idea from the papers \cite{CRS1,CR,CRS2} where,
instead of having a fixed number of excitations, we only fix their maximal number $N$.

Before giving the explicit form of the Bethe eigenvectors, we need some definitions:
\ben
\cI_\ell^N &=& \{\  \{i_1,i_2,\dots,i_\ell\} \ |\ 1\leq i_{1} <  \dots <  i_{\ell}\leq N\ \}\,, \quad0\leq \ell\leq N\,,
\\
\cI^N &=& \bigcup_{\ell=0}^N\, \cI_\ell^N,
\\
\overline I&=&\{1,2,\dots,N\}\setminus I \,,\qquad \forall\, I\in \cI^N.
\een
Starting from a set of $N$ elements $\mathbf{u}=\{ u_1,\dots,u_N \}$, we define $\mathbf{u}_I=\{u_i\ |\ i\in I\}$ for any 
$I\in\cI^N$.

We can now define the Bethe vectors which depend on the Bethe parameters $\mathbf{u}=\{u_1, u_2,\dots u_N\}$:
\ben\label{BV}
|\Phi^N(\mathbf u)\rangle&=&\sum_{I\in \cI^N} \
W_{I}\left(\mathbf{u}\right)\ |\BB(\mathbf{u}_{I})\rangle.
\een
We have used the following vectors:
\ben
|\BB(\mathbf{u}_{I})\rangle=\prod_{i\in I} \mB(u_i)|\Omega\rangle\,,\qquad I\in \cI^N
 \een
and functions
\ben
W_I\left(\mathbf{u}\right)&=&
\frac{\displaystyle \prod_{i\in \overline I}\left(\Lambda_2(u_i)\ \frac{\bar{c} (2 u_i+\eta)}{2(\bar{b} u_i-\bar{a})}
\prod_{k=1\atop k\neq i}^N h(u_i,u_k) \right)}
{\displaystyle  \prod_{j,k \in \overline I \atop j<k} h(u_j,u_k) f(u_j,u_k) }\, ,\qquad I\in \cI^N,
\label{def:W}
\een
where  $f$ and $h$ are defined by
\ben
\label{eq:fg}
f(u,v)= \frac{(u-v-\eta)(u+v)}{(u+v+\eta)(u-v)} \mb{and}  h(u,v)= \frac{(u-v+\eta)(u+v+2\eta)}{(u-v)(u+v+\eta)} \,. 
\een

Before giving the main result of this paper, let us comment the form of the Bethe vectors to clarify notations and  conventions.
We use the usual convention that a product over an empty set is equal to one, \textit{e.g.} if $I\in \cI_N^N$, we get $\overline{I}=\emptyset$ and $W_I=1$.
We deduce that, for $\bar{c}=0$, we get
\ben\label{BVco}
|\Phi^N(\mathbf u)\rangle&=&\prod_{i=1}^N \mB(u_i)|\Omega\rangle.
\een
It corresponds to the usual form taken by the ansatz for a diagonal left boundary, since for $\bar{c}=0$ the left boundary is indeed diagonal and the operator $\mC$ is not present 
anymore in the transfer matrix (\ref{eq:tt}).
For $N=0$ (and any $\bar c$), the Bethe vector $|\Phi^0\rangle$ reduces to the pseudo-vacuum $|\Omega\rangle$.

\subsection{Eigenvectors and eigenvalues\label{sec:EE}}

Let us now present the main result of this paper, whose proof is given in the next section. The Bethe vector (\ref{BV})  is an eigenvector of the transfer matrix $\ft(u)$ 
defined by (\ref{eq:tt}) i.e. we get
\ben\label{eq:vpb}
\ft(u)|\Phi^N(\mathbf u)\rangle=\Lambda(u)  |\Phi^N(\mathbf u)\rangle
\een
with
\ben\label{eigenval}
\Lambda(u)=
\ck_1(u) \Lambda_1(u)\ \prod_{k=1}^{N} f(u,u_k)  
+\ck_2(u)  \Lambda_2(u)\ \prod_{k=1}^{N} h(u,u_k)
\een
if the Bethe parameters $\{u_1,u_2,\dots,u_N\}$,  that are supposed to be all different, satisfy the following Bethe equations
\beq\label{eq:bethe}
\frac{\Lambda_1(u_k)}{\Lambda_2(u_k)}=\Xi(u_k)\prod_{j\neq k}^{N}\frac{h(u_k,u_j)}{f(u_k,u_j)}
\eeq
with
\ben
\Xi(u)=\frac{(2\,u+\eta)( \bar{b}(u+\eta)+\bar{a})}{2\,u (\bar{a}-\bar{b}u)}.
\een

\section{Proof of relation (\ref{eq:vpb})\label{sec:pr}}

\subsection{Actions of $\mA(u),  \mD(u)$ and $\mC(u)$}

We need the actions of $\mA(u)$, $\mD(u)$ and $\mC(u)$ on vectors of 
 type $|\BB(\mathbf{x})\rangle=\prod_{i=1}^\ell \mB(x_i)|\Omega\rangle$ 
where $\mathbf{x}=\{ x_1,\dots,x_\ell \}$ can be any subset of Bethe parameters.
The computation for the first two actions is well known \cite{sklyanin}  and based on the commutation relations of 
 Appendix \ref{App:cr}. We get
\ben
 \mA(u) |\BB(\mathbf{x})\rangle  &=&  
 \Lambda^\ell_1(u,\mathbf{x})   |\BB(\mathbf{x})\rangle 
 + \sum_{k=1}^{\ell} \, M_k(u,\mathbf{x})\ \mB(u)\ |\BB(\mathbf{x}_{\neq k})\rangle\,,
\label{AB}\\
 \mD(u) |\BB(\mathbf{x})\rangle  &=&  
 \Lambda^\ell_2(u,\mathbf{x})   |\BB(\mathbf{x})\rangle 
 + \sum_{k=1}^{\ell} \, N_k(u,\mathbf{x})\  \mB(u)\ |\BB(\mathbf{x}_{\neq k})\rangle\,,
\label{DB}
\een
where $\mathbf{x}_{\neq k}=\{ x_1,\dots,x_{k-1},x_{k+1},\dots,x_\ell \}$ and
\ben\label{eq:Ll}
\Lambda_1^\ell(u,\mathbf{x})&=&\Lambda_1(u)\ \prod_{k=1}^{\ell} f(u,x_k)   \mb{,}
\Lambda_2^\ell(u,\mathbf{x})=\Lambda_2(u)\ \prod_{k=1}^{\ell}  h(u,x_k) \,,
\\
\label{eq:M}
M_k(u,\mathbf{x})&=& g(u,x_k)\ \Lambda_1^{\ell-1}(x_k\,,\mathbf{x}_{\neq k})+ w(u,x_k)\ \Lambda_2^{\ell-1}(x_k\,,\mathbf{x}_{\neq k})\,,\\
\label{eq:N}
N_k(u,\mathbf{x})&=& k(u,x_k)\ \Lambda_2^{\ell-1}(x_k\,,\mathbf{x}_{\neq k})+ n(u,x_k)\ \Lambda_1^{\ell-1}(x_k\,,\mathbf{x}_{\neq k}).
 \een

The computation of the action of $\mC(u)$ on $|\BB(\mathbf{x})\rangle$ is  more involved, but after some algebra, 
using the commutation relations of  Appendix \ref{App:cr}, we get
\ben\label{CB}
\mC(u) |\BB(\mathbf{x})\rangle &=&
\sum_{i=1}^\ell G_i(u,\mathbf{x})\ |\BB(\mathbf{x}_{\neq i})\rangle
+\sum_{i < j}^\ell
F_{ij}(u,\mathbf{x})\ \mB(u)\ |\BB(\mathbf{x}_{\neq i,j})\rangle\,,
\een 
 where $\mathbf{x}_{\neq i,j}=\{ x_1,x_2,\dots,x_\ell \}\setminus\{x_i,x_j\}$.
The functions $F$ and $G$ are given in terms of the functions defined in (\ref{eq:mz}) and by the following relations:
\ben\label{eq:defG}
G_i(u,\mathbf{x})&=&\Lambda^{\ell-1}_1(u,\mathbf{x}_{\neq i})
\Big(
\big(m(u,x_i)+l(u,x_i)\big)\Lambda^{\ell-1}_1(x_i,\mathbf{x}_{\neq i})
+p(u,x_i)\Lambda^{\ell-1}_2(x_i,\mathbf{x}_{\neq i})\Big)\nonumber\\
&+&\Lambda^{\ell-1}_2(u,\mathbf{x}_{\neq i})
\Big(\big(q(u,x_i)+y(u,x_i)\big)\Lambda^{\ell-1}_1(x_i,\mathbf{x}_{\neq i})
+z(u,x_i)\Lambda^{\ell-1}_2(x_i,\mathbf{x}_{\neq i})\Big)\,,\\
F_{ij}(u,\mathbf{x})&=&
\Lambda^{\ell-2}_1(x_i,\mathbf{x}_{\neq i,j})
\left(Z_{11}(u,x_i,x_j)\Lambda^{\ell-2}_1(x_j,\mathbf{x}_{\neq i,j})
+Z_{12}(u,x_i,x_j)\Lambda^{\ell-2}_2(x_j,\mathbf{x}_{\neq i,j})\right)\nonumber\\
&+&
\Lambda^{\ell-2}_2(x_i,\mathbf{x}_{\neq i,j})
\left(
Z_{12}(u,x_j,x_i)\Lambda^{\ell-2}_1(x_j,\mathbf{x}_{\neq i,j})
+Z_{22}(u,x_i,x_j)\Lambda^{\ell-2}_2(x_j,\mathbf{x}_{\neq i,j})\right)\,,\qquad
\label{eq:F}
\een  
where
\ben
Z_{11}(u,x_i,x_j)&=&\frac{8\eta^2x_ix_j(x_i+x_j)(u^2-x_ix_j+\eta u)}{(2x_i+\eta)(2x_j+\eta)(x_i+x_j+\eta)(u+x_i+\eta)(u+x_j+\eta)(u-x_i)(u-x_j)},\quad\ \\
Z_{12}(u,x_i,x_j)&=&\frac{4\eta^2 x_i(x_j-x_i+\eta)(u^2+\eta u +x_ix_j +\eta x_i)}{(2x_i+\eta)(x_i-x_j)(u+x_i+\eta)(u+x_j+\eta)(u-x_i)(u-x_j)},\\
Z_{22}(u,x_i,x_j)&=&\frac{2\eta^2(x_i+x_j+2\eta)(u^2-(x_i+\eta)(x_j+\eta)+\eta u)}{(x_i+x_j+\eta)(u+x_i+\eta)(u+x_j+\eta)(u-x_i)(u-x_j)}.
\een

 A direct calculation, using relations (\ref{AB}), (\ref{DB}) and (\ref{CB}), shows that
 \begin{eqnarray}\label{eq:vpb2}
&&\!\! \big(\ft(u)-\Lambda(u)\big) |\Phi^N(\mathbf u)\rangle =
 \sum_{\ell=0}^N \sum_{I\in I_\ell^N} W_{I}(\mathbf{u})
 \Big\{
 \Big(\ck_1(u)\Lambda_1^\ell(u,\mathbf{u}_I)+\ck_2(u)\Lambda_2^\ell(u,\mathbf{u}_I)-\Lambda(u)\Big)|\BB(\mathbf{u}_{I})\rangle\nonumber\\
 &&\qquad\qquad\qquad
 +\sum_{i\in I}\Big[\Big(\ck_1(u) M_i(u,\mathbf{u}_{I})+ \ck_2(u) N_i(u,\mathbf{u}_{I})\Big)\mB(u)
 +  \ck_{12}(u) G_i(u,\mathbf{u}_{I}) \Big] |\BB(\mathbf{u}_{I\backslash i})\rangle\nonumber\\
&&\qquad\qquad\qquad
+\ck_{12}(u)\sum_{i,j\in I,i<j} F_{i,j}(u,\mathbf{u}_{I})\mB(u)|\BB(\mathbf{u}_{I\backslash \{i,j\}})\rangle\Big\}\,.
 \end{eqnarray}
Then, proving (\ref{eq:vpb}) amounts to show that in (\ref{eq:vpb2}), the coefficients associated with all vectors $|\BB(\mathbf{u}_{I})\rangle$, 
 $I\in \cI^N$,  and all vectors
 $\mB(u)|\BB(\mathbf{u}_{I})\rangle$, $I\in \cI^N$ vanish. The following sections deal with these different coefficients.

\subsection{Coefficient of $|\BB(\mathbf{u})\rangle$}

This computation is similar to the one done for diagonal boundaries. Thus, we just sketch the proof.
The only contributions   to this coefficient come from the first line of 
  (\ref{eq:vpb2})  for $\ell=N$. Using the explicit expressions (\ref{eq:Ll}), it corresponds to
the  form (\ref{eigenval}) for the eigenvalue.
This makes the coefficient of  $|\BB(\mathbf{u})\rangle$ in (\ref{eq:vpb2}) vanish.

\subsection{Coefficient of $\mB(u)|\BB(\mathbf{u}_{\neq k})\rangle$}

This computation is also similar to  the usual one with diagonal boundaries.
The coefficient takes the following form:
\begin{equation}
\label{eq:probet}
 \ck_1(u)M_k(u,\mathbf{u})+\ck_2(u)N_k(u,\mathbf{u}).
\end{equation}
By applying Bethe equations (\ref{eq:bethe}) and using the 
explicit forms (\ref{eq:M}) and (\ref{eq:N}), it follows that (\ref{eq:probet}) is equal to zero.

\subsection{Coefficient of $|\BB(\mathbf{u}_{I})\rangle$ for $I\in \cI_{\ell}^{N}$ and $\ell=0,1,\dots,N-1$}

This type of vector is not present in the usual diagonal boundary case. Therefore, we give here more details.
The coefficient that should be zero is
\ben
W_{I}(\mathbf{u})\Big(\ck_1(u)\Lambda_1^\ell(u,\mathbf{u}_I)+\ck_2(u)\Lambda_2^\ell(u,\mathbf{u}_I)-\Lambda(u)\Big)
+\ck_{12}(u)\sum_{j\in \overline{I}}W_{I\cup j}(\mathbf{u})\ G_j(u,\mathbf{u}_{I\cup j})\,,
\label{eq1}
\een
 where $\mathbf{u}_{I\cup j}=\{u_k\}_{k\in I}\cup\{u_j\}$.
In fact we are going to prove a stronger statement: the coefficient of 
$\Lambda^\ell_1(u,\mathbf{u}_I)$ in (\ref{eq1}), as well as the coefficient of $\Lambda^\ell_2(u,\mathbf{u}_I)$ in the same equation, both vanish. Indeed, the first coefficient reads
\ben
\c_1(u) &=& \ck_1(u)W_{I}(\mathbf{u})\big(1-\prod_{j\in \overline{I}} f(u,u_j)\big)
\nonu
&&+\ck_{12}(u)\sum_{j\in \overline{I}}W_{I\cup j}(\mathbf{u})
\Big(
\big(m(u,u_j)+l(u,u_j)\big)\Lambda^{\ell}_1(u_j,\mathbf{u}_{I})
+p(u,u_j)\Lambda^{\ell}_2(u_j,\mathbf{u}_{I})\Big)
\een
while the second one is just $\c_2(u)=\c_1(-u-\eta)$. Thus, if one coefficient vanishes for all $u$, so does the other one. 

To show that $\c_1(u)$ vanishes, we follow the  technics used in \cite{CRS1,CR}: we prove that $\c_1(u)$ corresponds to the sum over all residues of some rational function. The function to consider is
\beq
F(z,u)=\frac{\bar{b} z-\bar{a}}{2\eta z (z-u)}\prod_{j \in \overline{I}} f(z,u_j).
\eeq
The poles of $F$ (considered as a rational function of $z$) are located at $z=u_j$, $-u_j-\eta$, 0, $u$ and $\infty$. 
It is a simple exercise to show that $\c_1(u)$ is the sum over all the residues of $F$, and therefore vanishes. To do this calculation, one has to replace
the functions $W_{I}$, $\Lambda^{\ell}_k$, $\ck_1$, $\ck_{12}$, $m$, $l$ and $p$ by 
their explicit expression \eqref{def:W}, (\ref{eq:Ll}), (\ref{eq:kappa}) and (\ref{eq:mz}), and
to use the Bethe equations (\ref{eq:bethe}).
This proves that the coefficients of $|\BB(\mathbf{u}_{I})\rangle$ for $I\in \cI_{\ell}^{N}$ and $\ell=0,1,\dots,N-1$ in  (\ref{eq:vpb2}) vanish.

\subsection{Coefficient of $\mB(u)|\BB(\mathbf{u}_{I})\rangle$ for $I\in \cI_{\ell}^{N}$ and $\ell=0,1,\dots,N-2$}

This term is also new in comparison with the case with diagonal boundaries. 
The coefficient is
\ben
X(u)&=&\sum_{j\in \overline{I}} 
W_{I\cup j}(\mathbf{u})\Big(\ck_1(u)M_j(u,\mathbf{u}_{I\cup j})+\ck_2(u)N_j(u,\mathbf{u}_{I\cup j})\Big)
\nonu
&&
+\ck_{12}(u)\sum_{j,k\in \overline{I} \atop j<k}W_{I\cup \{j,k\}}(\mathbf{u}) F_{ij}(u,\mathbf{u}_{I\cup \{j,k\}})\,.
\label{eq2}
\een
Using the explicit expressions 
of $W_{I}$, $M_j$, $N_j$, $F_{ij}$ $\ck_1$, $\ck_{12}$ 
(see relations \eqref{def:W}, (\ref{eq:M}), (\ref{eq:N}), (\ref{eq:F}) and (\ref{eq:kappa})) and
 the Bethe equations (\ref{eq:bethe}), we can  rewrite $X(u)$ as 
\begin{eqnarray}
\label{eq:pp}
X(u)&=&\sum_{j\in \overline{I}}L(u,u_j)
\Big(\prod_{k\in \overline{I}\atop k\neq j }h_{jk}-\prod_{k\in \overline{I}\atop k\neq j }f_{jk}
\Big)+\frac{1}{2}\sum_{j,\ell\in \overline{I} \atop j\neq \ell}
\Big(Q(u,-u_j-\eta,-u_\ell-\eta)\prod_{k\in \overline{I} \atop k\neq j,\ell}f_{j k}f_{\ell k}+\\
&+&\!\!
Q(u,u_j,u_\ell)\prod_{k\in \overline{I} \atop k\neq j,\ell}h_{j k}h_{\ell k}
+Q(u,-u_j-\eta,u_\ell)\prod_{k\in \overline{I} \atop k\neq j,\ell}f_{j k}h_{\ell k}
+Q(u,u_j,-u_\ell-\eta)\prod_{k\in \overline{I} \atop k\neq j,\ell}h_{j k}f_{\ell k}\nonumber
\Big),
\end{eqnarray}
where $h_{jk}$ and $f_{jk}$ stand for $h(u_j,u_k)$ and $f(u_j,u_k)$ and
\begin{eqnarray}
 L(u,u_j)&=&2\ \big(\ck_1(u)g(u,u_j)+\ck_2(u)n(u,u_j)\big)\ \Xi(u_j)\ 
\frac{\bar{b}u_j-\bar{a}}{(2u_j+\eta)(u_j+\eta)}\,, \\
 Q(u,u_j,u_\ell)&=&-4\ Z_{11}(u,u_j,u_\ell)\ \Xi(u_j)\Xi(u_\ell)\ 
\frac{(\bar{b}u_j-\bar{a})(\bar{b}u_\ell-\bar{a})(u_j+u_\ell+2\eta)}
{(2u_j+\eta)(2u_\ell+\eta)(u_j+u_\ell)}.
\end{eqnarray}
It is easy to see that $X(u)$ is a rational function that tends to 0 when $u\rightarrow\infty$. 
We also remark that $X(u)$ can possess poles only at $u=u_j$ and $u=-u_j-\eta$.
Then, its residue at $u=u_j$ is equivalent to the sum over all the residues 
of the following rational function of $z$
\begin{equation}
 \frac{\bar{a}-\bar{b} z}{(z+u_j)(z-u_j-\eta)z}\prod_{\ell\in\overline{I}}f(z,u_\ell)
\end{equation}
and, by consequence, vanishes. We perform the same type of computation for the pole $u=-u_j-\eta$.
This makes  $X(u)$  a rational function that vanishes at infinity and has no pole:
it is  equal to 0.
This proves that the coefficient of $\mB(u)|\BB(\mathbf{u}_{I})\rangle$ for $I\in \cI_{\ell}^{N}$ and $\ell=0,1,\dots,N-2$ in  (\ref{eq:vpb2}) vanishes, and concludes 
the proof of relation (\ref{eq:vpb}).

\section{Conclusion}
We have constructed the algebraic Bethe ansatz for the XXX model with open boundary conditions characterized by 
two general boundary matrices related by one constraint. This constraint means that the two boundary matrices can be triangularized in the same basis. 
This new method should be put in correspondence with the Vertex-IRF correspondence used in the XXZ model \cite{CLSW} and 
allowed only when constraints are applied to the boundary matrices. We believe that the present method is much simpler than the one of \cite{CLSW}
and can give a better insight into the problem of non-diagonal boundaries. 
However, the direct comparison between the two approaches is rather difficult at this point, since the one presented here deals with the XXX model, while the Vertex-IRF correspondence is done for the XXZ model: a careful limit of the Vertex-IRF correspondence, or the generalization to the XXZ model of the present approach, should be done to clarify this point.

Several directions of investigation can follow. Keeping the same model, one should 
compute the scalar products of (off-shell) Bethe vectors to have access to the correlation functions 
of the model. The same construction for the XXZ model with non-diagonal boundaries should 
also be  done and compared with the previous results \cite{nepo,CLSW,BK,niccoli,niccoli2}. 
Generalization to different models, such as spin chains based on algebras of higher rank or the Hubbard model, should be also investigated.
Finally, the seeking of an algebraic Bethe ansatz for general boundary matrices 
is a very exciting, but still unresolved, problem.

\vspace{0.1cm}

{
\noindent{\bf Acknowledgements:}  The authors thank the referees for their constructive remarks.}

\appendix
\section{Commutation relations\label{App:cr}}

Using the reflection equation (\ref{RA}), we can find the exchange relations between the operators $\mA$,
$\mB$, $\mC$ and $\mD$.
For our calculations, we only need  the following ones
\ben
[ \mB(u) , \mB(v) ]&=& 0\,, \\
\mA(u)\mB(v)&=& f(u,v)\mB(v)\mA(u) + g(u,v)\mB(u)\mA(v) + w(u,v)\mB(u)\mD(v),\nonumber \\
\mD(u)\mB(v)&=& h(u,v)\mB(v) \mD(u) +k(u,v)\mB(u)\mD(v)+ n(u,v)\mB(u)\mA(v)\,,\nonumber\\
\null[\mC(u),\mB(v)]&=&m(u,v) \mA(v) \mA(u)+l(u,v) \mA(u) \mA(v)+q(u,v) \mA(v)\mD(u)\\
&&+p(u,v)\mA(u)\mD(v)+y(u,v) \mD(u) \mA(v)+z(u,v) \mD(u) \mD(v)\,,\nonumber
\label{ExchDB}
\een
with $f$ and $g$ given in (\ref{eq:fg}),
\ben
\label{eq:fh}
&&w(u,v)= \frac{-\eta}{(u+v+\eta)}, \quad  g(u,v)= \frac{2\,\eta\,v}{(2v+\eta)(u-v)} \,,  \\
&&  k(u,v)= \frac{-2\,\eta\,(u+\eta)}{(u-v)(2u+\eta)} , \quad
n(u,v)= \frac{4\,v\,\eta\,(u+\eta)}{(u+v+\eta)(2v+\eta)(2u+\eta)}\,. \nonumber
\een
and
\ben
\label{eq:mz}
m(u,v)&=&\frac{2\, \eta\, u\,(u-v+\eta)}{(2\,u+\eta)(u+v+\eta)(u-v)},\quad
l(u,v)=-\frac{2\, \eta^2 \,u}{(2\,u+\eta)(2\,v+\eta)(u-v)}\,,\\
q(u,v)&=&\frac{\eta\,(u+v)}{(u+v+\eta)(u-v)} ,\quad
p(u,v)=-\frac{2\, \eta \,u}{(2\,u+\eta)(u-v)}\,,\nonumber\\
y(u,v)&=&-\frac{\eta^2}{(u+v+\eta)(2v+\eta)} ,\quad 
z(u,v)=-\frac{\eta}{u+v+\eta}.
\nonumber
\een


\begin{thebibliography}{99}

\bibitem{xxz} Alcaraz, F., Barber, M., Batchelor, M., Baxter, R. and Quispel, G.: 
\textsl{Surface exponents of the quantum XXZ, Ashkin-Teller and Potts models}, 
J. Phys. \textbf{A20} (1987) 6397.
  
\bibitem{Gb}
Arnaudon, D., Avan, J., Cramp\'e, N., Doikou, A., Frappat, L. and Ragoucy, E.:
\textsl{General boundary conditions for the $sl(N)$ and $sl(M|N)$ open spin chains,}
J. Stat. Mech. 0408 P08005 (2004) and \texttt{arXiv:math-ph/0406021}

\bibitem{BK} Baseilhac, P.  and Koizumi, K.: 
\textsl{Exact spectrum of the XXZ open spin chain from the q-Onsager
algebra representation theory}, JSTAT 0709 P09006 (2007) and \texttt{arXiv:hep-th/0703106}.

\bibitem{baxter}
Baxter, R.J.: \textsl{Partition function of the eight-vertex lattice model,}  Ann. Phys. 70 193-228 (1972).

\bibitem{baxter2}
Baxter, R.J.:  \textsl{Asymptotically degenerate maximum eigenvalues of the eight-vertex 
  model transfer matrix and interfacial tension},
J. Stat. Phys. 8 25-55 (1973).

\bibitem{baxter3}
Baxter, R.J., \textsl{Exactly solved models in statistical mechanics}, Academic 
Press, London (1982).

\bibitem{OBA} 
Belliard, S. and Ragoucy, E.:
\textsl{The nested Bethe ansatz for 'all' open spin chains
with diagonal boundary conditions}, 
J. Phys. A 42 205203 (2009) and \texttt{arXiv:0902.0321}.

\bibitem{bethe}
Bethe, H.: \textsl{Zur Theorie der Metalle. Eigenwerte und Eigenfunktionen 
Atomkete,} Zeitschrift f{\"u}r Physik 71  205-226 (1931).

\bibitem{CLSW}
Cao, J., Lin, H., Shi, K. and Wang, Y.:
\textsl{Exact solutions and elementary excitations in
the XXZ spin chain with unparallel boundary fields}, 
Nucl. Phys. B 663 487-519 (2003) and \texttt{arXiv:cond-mat/0212163}.

\bibitem{chered}
Cherednik, I.V.: 
\textsl{Factorizing particles on a half line and root systems}, 
Theor. Math. Phys. 61 977-983 (1984).

\bibitem{CR} Cramp\'e, N.  and Ragoucy, E.:
\textsl{Generalized coordinate Bethe ansatz for non diagonal boundaries,}
Nucl. Phys. B 858 502-512 (2012) and \texttt{arXiv:1105.0338}.  

\bibitem{CRS1} Cramp\'e, N., Ragoucy, E. and Simon, D.:
\textsl{Eigenvectors of open XXZ and ASEP models for a class of non-diagonal boundary conditions,}
J. Stat. Mech. P11038 (2010) and \texttt{arXiv:1009.4119}. 

\bibitem{CRS2} Cramp\'e, N., Ragoucy, E. and Simon, D.:
\textsl{Matrix Coordinate Bethe Ansatz: Applications to XXZ and ASEP models,}
J. Phys. A 44 405003 (2011) and \texttt{arXiv:1106.4712}.

\bibitem{Kmat} 
de Vega, H.  and Gonzalez-Ruiz, A.:
\textsl{Boundary K-matrices for the six vertex and the n(2n-1) $A_{n-1}$ vertex models}, 
J. Phys. A 26 L519-L524 (1993)  and \texttt{arXiv:hep-th/9211114}.

\bibitem{fafa} Faddeev, L.D.: \textsl{How Algebraic Bethe Ansatz works
for integrable model}. In \textit{Sym\'etries Quantiques},
Les Houches summer-school proceedings 64, Eds A. Connes, K. 
Gawedzki
and J. Zinn-Justin, North-Holland 1998, and
\texttt{hep-th/9605187}.

\bibitem{FSW} Frahm, H., Seel, A. and Wirth, T.: \textsl{Separation of Variables in the open XXX chain}, Nucl. Phys. B 802 351-367 (2008)  and \texttt{arXiv:0803.1776}.

\bibitem{GM} Galleas, W.  and Martins, M.J.: 
\textsl{Solution of the SU(N) Vertex Model with Non-Diagonal Open Boundaries},
Phys. Lett. A 335 167-174 (2005)  and \texttt{arXiv:nlin/0407027}.

\bibitem{gaudin} Gaudin, M.: \textsl{Boundary Energy of a Bose Gas in One Dimension}, Phys. Rev. 4 386-394 (1971).

\bibitem{heisen}
Heisenberg, W.: \textsl{Zur Theorie der Ferromagnetismus}, Zeitschrift f{\"u}r Physik 49 619-636 (1928).

\bibitem{KoIzBo}
Korepin, V.E., Izergin, G. and Bogoliubov, N.M.:
  \textsl{Quantum inverse
    scattering method, correlation functions and algebraic Bethe 
Ansatz}, Cambridge University Press, Cambridge, (1993).

\bibitem{kusk} Kulish, P.P. and Sklyanin, E.K.:   \textsl{Quantum inverse scattering method and the Heisenberg ferromagnet}, Phys. Lett. A 70 461-463 (1979). 

\bibitem{MMR} Melo, C.S., Martins,  M.J.  and Ribeiro, G.A.P.:  \textsl{ Bethe ansatz for the XXX-S chain with non-diagonal open boundaries},
Nucl. Phys. B 711 565-603 (2005) and \texttt{arXiv:nlin/0411038}.

\bibitem{MTV} Mukhin, E. ,Tarasov, V. and, Varchenko A.: Bethe algebra of homogeneous XXX Heisenberg model has simple spectrum, Comm. Math. Phys. 288 1-39 (2009) and \texttt{arXiv:0706.0688}.

\bibitem{nepo}
Nepomechie, R.I.:  \textsl{Bethe Ansatz solution of the open XXZ chain with nondiagonal boundary terms}, J. Phys. A 37 433-440 (2004) and arXiv:hep-th/0304092;

\bibitem{niccoli} Niccoli, G.:  \textsl{Antiperiodic spin-1/2 XXZ quantum chains by separation of variables: Complete spectrum and form factors}, \texttt{arXiv:1205.4537}.

\bibitem{niccoli2}Niccoli, G.:  \textsl{Non-diagonal open spin-1/2 XXZ quantum chains by separation of variables: Complete spectrum and matrix elements of some quasi-local operators}, \texttt{arXiv:1206.0646}.

\bibitem{sklyanin}
Sklyanin, E.K.:  \textsl{Boundary conditions for integrable quantum 
systems}, J. Phys. A 21 2375-2389 (1988).

\bibitem{tafa}Takhtajan, L.A. and Faddeev, L.D.:  \textsl{The Quantum method of the inverse problem and the Heisenberg XYZ model}, Russ. Math. Surveys 34 11-68 (1979).

\bibitem{yang}
 Yang, C.N.:  \textsl{Some exact results for the many-body problem in 
one dimension with repulsive delta-function interaction}, Phys. Rev. Lett. 19 1312-1315 (1967).
\end{thebibliography}
\end{document}